\documentclass[twocolumn,final,nofootinbib]{elsarticle}

\usepackage{amsmath,amssymb}
\usepackage{lipsum}
\usepackage{amsthm}
\usepackage{graphicx}
\usepackage{hyperref}
\usepackage{booktabs}
\journal{Physics Letters B}
\begin{document}
\begin{frontmatter}
\title{Higgs Field as Architect of a Geodesically Complete Universe\\and Agent for New Physics in Interiors of Black Holes}
\author{Itzhak Bars}
\address{Department of Physics and Astronomy, University of Southern California, Los Angeles, CA 90089-0484, USA}
\begin{abstract}
The Higgs boson is often compared to a small ripple on a vast, calm ocean.
This \textquotedblleft ocean\textquotedblright\ is the universal Higgs vacuum
expectation value, $\approx246$ GeV, permeating all space and generating
particle masses. Its remarkable uniformity reflects the large-scale nature of
the universe. Here I argue that the Higgs field plays a far deeper role. In
regions of extreme gravity near singularities, it becomes a non-perturbative,
space-time--dependent field that creates new space-time domains beyond
gravitational singularities---new regions governed by antigravity. This
mechanism produces dramatic dynamics inside black holes and in the earliest
universe, including restoration of the electroweak symmetry SU(2)$\times$U(1)
at singularities. The adjoining gravity and antigravity regions form a
geodesically complete space-time bridged by traversable singularities---absent
in the Standard Model coupled to general relativity or its extensions such as
string theory. In such a framework, the black hole information paradox admits
a new, manifestly unitary resolution. Details of computations are developed in
the companion paper \cite{IBbhHinterior}; here, only a brief outline is given.
\end{abstract}
\begin{keyword}
Higgs Field, Black Hole, Standard Model, General Relativity,
antigravity, conformal symmetry, geodesic completeness, Kruskal
diagram, Penrose diagram
\end{keyword}
\end{frontmatter}
%
%

\section{A fundamental problem and its resolution}

The Standard Model (SM) with General Relativity (GR) is
\textit{geodesically\ incomplete}, a problem persisting in practically all
quantum gravity approaches.

For a particle falling into a black hole, an external observer records in
spherical coordinates $(t(\tau),r(\tau))$ as functions of proper time $\tau$.
At the horizon $t\left(  \tau\right)  $ blows up, so the external observer
cannot keep track of the particle, but the particle does move to the center
$r=0$ in finite proper time $\tau$. The theory gives no guidance on what
happens at the next instant in proper time---true for all spacetime
singularities. This is the incompleteness of the (SM+GR) in $\left(
t,r\right)  $ spacetime.

In maximally extended Kruskal--Szekeres $(u,v)$ spacetime \cite{MTW}, physical
regions $I-IV$ satisfy $uv<1$, while $V,VI$ with $uv>1$ are discarded (since
$r=|\vec{r}|$ is strictly positive). A geodesic from region $I$ passes into
$II$, ending at $uv=1$ ($r=0$), with no prescription about the physics beyond.
This is the incompleteness in $\left(  u,v\right)  $ spacetime.

The resolution of this incompleteness is often deferred to quantum gravity,
but string theory, and other quantum gravity formalisms, introduce background
fields that are themselves incomplete, while non-geometric string models (e.g.
matrix models \cite{matrix}) obscure and don't address the incompleteness.

The root cause of incompleteness in both field theory and string theory is
deeper. It is the absence of \textit{antigravity domains} ($G<0$) in the
traditional formulation of classical or quantum gravity. In the presence of
additional antigravity spacetime domains, a geodesically complete universe is
achieved naturally, as discussed below.

A field theory resolution of completeness proposed in \cite{BST}, and later
extended to string theory \cite{BSTstrings}, invokes \textit{local scale
invariance} in an improved version of SM+GR. The field theory action is,
$S_{i(\mathrm{SM+GR})}=\int d^{4}x\,\sqrt{-g}\,L(x)$,%
\begin{equation}
L=\left(
\begin{array}
[c]{c}%
L_{SM}+\frac{1}{12}\left(  \phi^{2}-2H^{\dagger}H\right)  R\left(  g\right)
-V\left(  \phi,H\right) \\
+\frac{1}{2}g^{\mu\nu}\left(  \partial_{\mu}\phi\partial_{\nu}\phi
-2\partial_{\mu}H^{\dagger}\partial_{\nu}H\right)
\end{array}
\right)  \label{ActioA2}%
\end{equation}
Here $L_{SM}\left(  \psi,A_{\mu},g_{\mu\nu},H\right)  $ is the usual
SM Lagrangian, with quarks, leptons, gauge bosons, $H$ is the Higgs
doublet, $\phi$ is an additional scalar singlet, while gravity
couples non-minimaly with the conformal coupling of scalars to the
curvature $R\left(  g\right)  ,$ with fixed coefficient $1/12$. This
structure is locally scale invariant \cite{IBbhHinterior}, but does
not include the Weyl vector \cite{Geometry}. It is conceptually
different, as the local scale transformation has a geometric origin
in general coordinate transformations in 4+2 dimensional in
2T-physics \cite{BST}. Dimensionful constants are forbidden; in
i(SM+GR), all familiar scales emerge from $\phi$.

The crucial difference with the traditional (SM+GR) is that the Newton
constant $G_{N}$ in the usual Einstein--Hilbert term $\left(  16\pi
G_{N}\right)  ^{-1}R$ is replaced in (\ref{ActioA2}) by a dynamical
gravitational strength $G\left(  x\right)  $ %

\begin{equation}
\left(  16\pi G\left(  x\right)  \right)  ^{-1}\equiv\phi^{2}
\frac{1-h^{2} }{12},\text{ }h^{2}\equiv\frac{2H^{\dagger}%
H}{\phi^{2}}\text{.} \label{G}%
\end{equation}
The factor $\left(  1-h^{2}\left(  x\right)  \right)  ,$ that is
local scale invariant, can change signs depending on location
$x^{\mu}$. The relative minus sign in (\ref{ActioA2}) for the
conformally coupled $\phi$ and $H$, in the curvature terms and in
the kinetic terms, is critical. It is required in order to have
regions of spacetime where the gravitational strength $G\left(
x\right)  $ is positive. Although $\phi$ appears ghost-like because
of the relative minus sign, local scale symmetry compensates the
ghost and preserves unitarity. \footnote{One way of arguing this
point is by noting that the ghost-like $\phi\left(  x\right)  $ can
be gauge fixed to a constant in each gravity/antigravity region of
spacetime, such as $\phi\left( x\right) =\phi_{+}\theta\left(
r\right) +\phi_{-}\theta\left( -r\right)  ,$ where the constants
$\phi_{\pm}$ in gravity/antigravity regions, are those exhibited in
the solution of the equations in Table.1 below. As already
mentioned, the gauge invariant $h\left(  x\right)  $ is not affected
by making such gauge choices. Since in such unitary gauges $\phi$ is
no longer a dynamical degree of freedom, the kinetic term for $\phi$
vanishes, so the theory is evidently unitary. Another convenient
gauge choice, called the "$\gamma$-gauge" in \cite{BST}, restricts
the metric by setting $\sqrt{-g}=1$ globally everywhere, in gravity
or antigravity, rather than restricting the scalar $\phi$ or some
convenient function of ($\phi,H$) in patches. Then the underlying
local conformal symmetry still must compensate for the ghost-like
$\phi$ by requiring unitarity in the quantization method as
discussed in detail in \cite{IBAJ}. In the latter gauge (as well as
other gauges), the issue is not unitarity, but rather how observers
in the gravity region can interpret certain phenomena that occur in
the antigravity region behind gravitational singularities. There are
no violations of any sacred principles, such as unitarity, but
instead there are opportunities to discuss various interesting
physical phenomena, in field theory and string theory \cite{IBAJ},
as perceived by observers such as us in the gravity region, within a
geodesically complete universe that includes regions of antigravity
behind gravitational singularities.}. This relative sign is the
gateway to new physics.

The potential satisfies $V(\Omega\phi,\Omega
H)=\Omega^{4}V(\phi,H)$, implying $V(\phi,H)=\phi^{4}v(h),$ with
$v(h)$ an arbitrary function of the scale-invariant ratio $h^{2}$ in
Eq. (\ref{G}).

At low energies---far from singularities---when $\phi\sim10^{19}$
$\mathrm{GeV}$ (Planck scale) and $\left\vert H\right\vert \sim246$
$\mathrm{GeV}$ (electroweak scale), the scale invariant $h\sim10^{-17}$ is
tiny. In this regime, choosing a local scale gauge where $\phi\left(
x\right)  $ is fixed to a constant ($\phi\left(  x\right)  \rightarrow\phi
_{0}\sim10^{19}$ $\mathrm{GeV}$) reproduces the standard model (SM+GR) because
the gravitational strength $G\left(  x\right)  $ reduces approximately to the
Newton constant: $\left(  16\pi G\left(  x\right)  \right)  ^{-1}\simeq
\frac{1}{12}\phi_{0}^{2}=\left(  16\pi G_{N}\right)  ^{-1}$, while $\phi$
ceases to be a field degree of freedom (no ghost). Moreover, the potential is
approximated by the renormalizable quartic form that fits observations:%

\begin{equation}%
\begin{array}
[c]{c}%
V(\phi,H)\rightarrow V_{4}\left(  \phi,s\right)  =\frac{\lambda}{4}\left(
s^{2}-\alpha^{2}\phi^{2}\right)  ^{2}+\frac{\lambda^{\prime}}{4}\phi^{4},\\
v\left(  h\right)  \rightarrow v_{4}\left(  h\right)  =\frac{\lambda}%
{4}\left(  h^{2}-\alpha^{2}\right)  ^{2}+\frac{\lambda^{\prime}}{4}.
\end{array}
\label{quartic}%
\end{equation}
Here $s$ is the Higgs field in the SU$\left(  2\right)  \times$U$\left(
1\right)  $ unitary gauge $H^{\dagger}=(0,s/\sqrt{2}),$ that satisfies
$s^{2}=2H^{\dagger}H.$ Under these conditions, \textit{i}(SM+GR) reproduces
identically the standard (SM+GR)'s phenomenology, including all the
dimensionful constants, namely the Newton's constant, the cosmological
constant, the Higgs vacuum expectation value, and the Higgs particle mass, all
emerging from the single scale $\phi_{0}$ \cite{BST}. The known phenomenology
fixes the parameters
\begin{equation}%
\begin{array}
[c]{c}%
\phi_{0}\approx0.596\times10^{19}~\text{GeV},\;\;\;\alpha\approx
4.13\times10^{-17},\\
\lambda^{\prime}\approx8.06\times10^{-122},\;\;\lambda\approx0.129.
\end{array}
\label{values2}%
\end{equation}
The tiny sizes of $\alpha$ and $\lambda^{\prime}/\lambda$ contribute to the
well known hierarchy puzzle that remains unresolved in $i$(SM+GR).

Thus, at low energies, $i(\mathrm{SM+GR})$ reproduces all tested SM+GR
predictions. Crucially, this well motivated theory is \textit{geodesically
complete}, permitting traversal of singularities---demonstrated in earlier
work for cosmology \cite{BSTloop}\cite{Sailing} and black holes \cite{BHoles},
and discussed further here and in \cite{IBbhHinterior}.

\section{Gravity and antigravity domains}

The dynamical gravitational coupling $G(x)$ is proportional to the factor
$\left(  1-h^{2}(x)\right)  .$ Thus $G(x)$ can be positive or negative in
different spacetime patches $x^{\mu}$, producing \textit{gravity} regions
($h^{2}(x)<1$) and \textit{antigravity} regions ($h^{2}(x)>1$) within the same
universe. Because $h$ is the ratio of two conformal fields, these domains are
scale gauge-invariant. This feature, absent in the traditional
$(\mathrm{SM+GR})$, is key to the \textit{geodesic} and \textit{field-space
completeness} of the improved theory $i(\mathrm{SM+GR})$.

The two types of regions meet at gravitational singularities where
$h^{2}(x)=1$. From Einstein's equations derived from
$i(\mathrm{SM+GR})$, $G_{\mu\nu}(x)=8\pi G(x)T_{\mu\nu}(x),$ one
sees that as $h^{2}(x)\rightarrow 1$, the gravitational strength
$G(x)$ diverges\footnote{When $G\left( x\right)  $ diverges near
singularities, the gravitational action becomes strongly coupled and
the treatment of the theory may be unreliable in that region. The
standard approach is to give up on the classical theory with the
hopes that some future quantum gravity theory will provide the
guidance. However, as stated in the introduction, and with more
detail in \cite{IBbhHinterior}, all available quantum gravity
formalisms are geodesically incomplete and fail to answer what
happens to a particle's trajectory that reaches the singularity. By
contrast, this paper argues that i(SM+GR) (and its string
generalization \cite{BSTstrings}) provides definite guidance to
where the particle or string goes in the geodesically complete
spacetime. The quantized string version in this approach is my
proposal for a geodesically complete quantum theory that could
reliably provide quantum effects that modify quantitatively the
physical phenomena revealed by the classical field theory.
Therefore, the classical i(SM+GR) is already valuable to begin this
process by revealing the first glimpses of new physical phenomena
that are unfamiliar in geodesically incomplete formalisms.}, driving
the geometry $G_{\mu\nu}(x)$ to a singularity. Such loci---black
holes, big bangs, etc.---are precisely the common spacetime
boundaries between gravity ($h^{2}<1$) and antigravity ($h^{2}>1$)
domains. I will argue that these gravity/antigravity regions are
unified into a geodesically complete spacetime via traversable
singularities located at $h^{2}\left(  x\right)  =1$.

Fig. 1 depicts the full scalar field space $(\phi(x^{\mu}),s(x^{\mu}))$. The
dashed diagonals mark singularities at $h^{2}(x)=1$ ($|\phi|=|s|$). The small
sphere at $+\infty$ represents a gravity-domain asymptotic region, far from
singularities, with $h\approx10^{-17}$ ($\phi\gg s$), corresponding to the
low-energy regime of our universe (see Table 1). The star at $-\infty$ marks
an asymptotic region in antigravity domain, where $h^{-1}$ is small
($|\phi|\ll|s|$). Three candidate curves connect these asymptotic regions,
representing hypothetical solutions $(\phi_{\mathrm{sol}}(x),s_{\mathrm{sol}%
}(x),g_{\mathrm{sol}}^{\mu\nu}(x))$ of the coupled scalar--geometry equations
derived from (\ref{ActioA2}). Each curve necessarily crosses the dotted lines
at a singularity $x_{0}^{\mu}$ where $\phi_{\mathrm{sol}}^{2}%
(0)=s_{\mathrm{sol}}^{2}(0)$. At this crossing: $\left(  \phi_{\mathrm{sol}%
}(0)+s_{\mathrm{sol}}(0)\right)  =0,$ while $\left(  \phi_{\mathrm{sol}%
}(0)-s_{\mathrm{sol}}(0)\right)  $ can be $(0,\ $finite$,\ \infty)$ for the
(red, black, blue) curves, respectively, as seen in Fig.1. Arrows indicate the
evolution of $(\phi_{\mathrm{sol}}\left(  x\right)  ,s_{\mathrm{sol}}\left(
x\right)  )$ as $x^{\mu}$ moves from $-\infty$ to $0$ to $+\infty$. The field
equations for $\left(  \phi\left(  x\right)  ,s\left(  x\right)  \right)  $
fix the full trajectory between asymptotic regions. It turns out that the
red-curve solution with $\phi_{\mathrm{sol}}(0)=s_{\mathrm{sol}}(0)=0$
dominates according to the analysis in \cite{IBbhHsolution}. This result
yields surprising insights into black-hole interiors, discussed below.%
\begin{figure}[htb]
\centering
        \includegraphics[width=0.7\columnwidth]{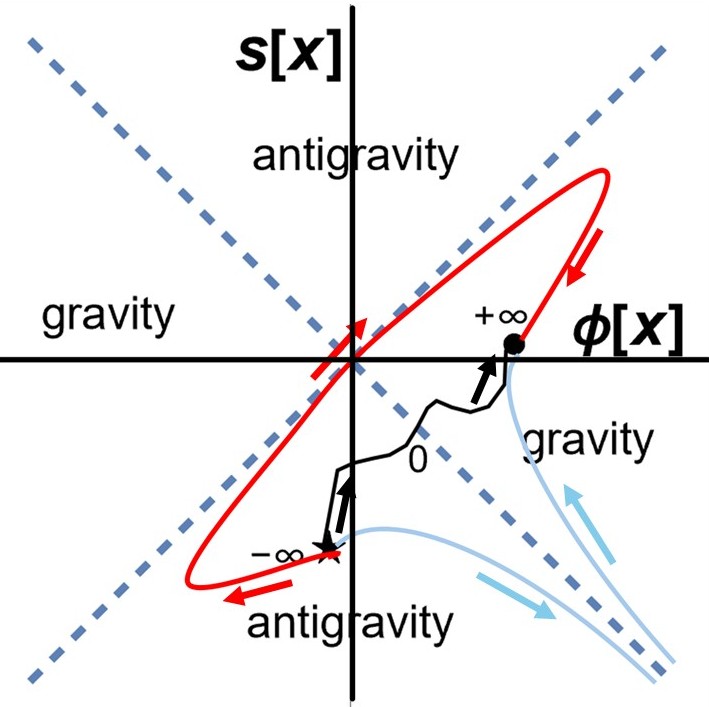}
        \caption{Scalar field space.}
        \label{fig:fig1}
\end{figure}
To describe both domains in one coordinate chart, choose local
systems $x_{\pm}^{\mu}=\left(  t_{\pm},\vec{r}_{\pm}\right)  $ in
gravity ($+$) and antigravity ($-$) regions, with the common
boundary at $h^{2}\left( x_{0}\right)  =1,$ located at
$\vec{r}_{+}=\vec{r}_{-}=0$ at any time $t_{+}=t_{-}=t,$ so that
$x_{0}^{\mu}\equiv(t,\vec{0})$. Dropping the $\pm$ labels for
convenience, define the new spherical coordinate symbol $r$ to mean
\begin{equation}
r\equiv\left\vert \vec{r}\right\vert \text{sign}\left(  1-h^{2}\left(
t,\vec{r}\right)  \right)  ,
\end{equation}
where, $\left\vert \vec{r}\right\vert =\sqrt{\vec{r}_{\pm}\cdot\vec{r}_{\pm}%
},$ is a positive distance as usual, but the new $r$ runs over the full range
$-\infty<r<\infty.$ Then, the vector $\vec{r}_{+}=\vec{r}$ is defined to be in
the gravity domain when $r>0,$ and the vector $\vec{r}_{-}=\vec{r}$ is defined
to be in the antigravity domain when $r<0$, while their shared boundary is at
the traversable singularity at $r=0$.

In these spherical coordinates ($t,r$), the Kruskal--Szekeres (KS) maximal
extension has the usual regions $I-IV$ ($r>0$) describing gravity. The
previously excluded $V,VI$ regions now correspond to $r<0$---not
\textquotedblleft negative distances\textquotedblright\ but genuine
antigravity domains $(t_{-},\vec{r}_{-})=(t,\vec{r}),$ where physical
phenomena occur.

In the KS diagram, gravity and antigravity $\left(  u,v\right)  $ regions are
distinguished, $\left(  I-IV\right)  $ with $uv<1$ versus $\left(
V,VI\right)  $ with $uv>1$, and are seamlessly joined at the singular $uv=1$
boundary. The fields live on a complete $\left(  u,v\right)  $ spacetime where
gravity and antigravity coexist as well-defined regions of the same universe
(for more clarity about KS diagrams, including the cosmological region, see
Figs. 9, 10 in \cite{IBbhHinterior}).

This geodesically complete structure, predicted by $i(\mathrm{SM+GR})$, is
absent in traditional (SM+GR) and extensions to string theory or other quantum
gravity approaches.

\section{Solutions of the field equations}

Consider first the gravity and antigravity asymptotic regions of a black hole,
denoted $\pm\infty$ in Fig.1. In these limits, all fields in $i(\mathrm{SM+GR}%
)$ vanish except $(\phi,s,g_{\mu\nu})$. The scalar fields approach constant
asymptotic values $(\phi_{\pm},s_{\pm})$, allowing us to drop their
derivatives. The curvature also approaches constant values $R_{\pm}$, so
derivatives of $g_{\mu\nu}$ are retained. Far from singularities, the scalar
potential $V(\phi,s)$ reduces to the purely quartic form $V_{4}(\phi,s)$ given
in Eq. (\ref{quartic}). The asymptotic field equations for $(\phi,s)$ are then%

\begin{equation}%
\begin{array}
[c]{c}%
\phi\left(  -\frac{R}{6}-\lambda\alpha^{2}\left(  s^{2}-\alpha^{2}\phi
^{2}\right)  +\lambda^{\prime}\phi^{2}\right)  =0,\\
s\left(  \frac{R}{6}+\lambda\left(  s^{2}-\alpha^{2}\phi^{2}\right)  \right)
=0.
\end{array}
\label{eomsConst}%
\end{equation}
Note the constant-$R$ terms absent in the traditional $(\mathrm{SM+GR})$
analysis. So, these equations are not obtained from minimizing only the
asymptotic potential energy $V_{4}\left(  \phi,s\right)  $. Define the Newton
constants in the asymptotic gravity $G_{N}$ and antigravity $\tilde{G}_{N}$
regions as%
\begin{equation}
\phi_{+}^{2}-s_{+}^{2}=\frac{+6}{8\pi G_{N}},\;\;\phi_{-}^{2}-s_{-}^{2}%
=\frac{-6}{8\pi\tilde{G}_{N}}, \label{gravOranti}%
\end{equation}
and insert these in the corresponding solutions $(\phi_{\pm},s_{\pm},R_{\pm})$
given in Table 1.
\begin{table}[h]
\centering \caption{Asymptotics in gravity/antigravity domains.}
\begin{tabular}{ll}
\toprule
gravity, $h^{2}<1$  &  antigravity, $h^{2}>1$ \\
\midrule
{$\phi_{+}^{2}=\frac{6}{8\pi G_{N}}\frac{1-\alpha^{2}%
}{\left(  1-\alpha^{2}\right)  ^{2}+\lambda^{\prime}/\lambda}$} &
{$\;\;\phi_{-}^{2}=0\text{ (or }small\text{)},\;$} \\
{$s_{+}^{2}=\frac{6}{8\pi
G_{N}}\frac{\alpha^{2}\left(1-\alpha^{2}\right)
-\lambda^{\prime}/\lambda}{\left( 1-\alpha^{2}\right)
^{2}+\lambda^{\prime}/\lambda}$} & {$\;\;s_{-}^{2}%
=\frac{6}{8\pi\tilde{G}_{N}}$} \\
{$h_{+}^{2}=\left( \alpha^{2}-\frac{\lambda^{\prime
}/\lambda}{1-\alpha^{2}}\right) \simeq\alpha^{2}$} &
{$\;\;h_{-}^{2}=\infty$ (or $large$)} \\
{$R_{+}=\frac{36}{8\pi G_{N}}\frac{\lambda^{\prime}%
}{\left(  1-\alpha^{2}\right)  ^{2}+\lambda^{\prime}/\lambda}$} &
{$\;\;R_{-}=-\frac{36\lambda}{8\pi\tilde{G}_{N}}$}\\
{$V_{4}\left(  \phi_{+},s_{+}\right) =\frac{R_{+}/4}{8\pi G_{N}}$} &
{$\;\;V_{4}\left( \phi_{-},s_{-}\right) =\frac{-R_{-}/4 }{
8\pi\tilde{G}_{N}}$}\\
{$\Lambda_{+}=R_{+}/4\simeq\frac{+9\lambda^{\prime}}{8\pi G_{N}}$} &
{$\;\;\Lambda_{-}=R_{-}/4\simeq\frac{-9\lambda
}{8\pi\tilde{G}_{N}}$}\\
\bottomrule
\end{tabular}
\label{tab:table1}
\end{table}
In the gravity domain, the smallness of $\alpha^{2}$ and $\lambda^{\prime
}/\lambda$ allows the approximations, $\phi_{+}^{2}\simeq\frac{6}{8\pi G_{N}%
},\quad s_{+}^{2}\simeq\frac{6\alpha^{2}}{8\pi G_{N}},\quad R_{+}\simeq
\frac{+36\lambda^{\prime}}{8\pi G_{N}}.$ In the antigravity domain, introduce
a dimensionless parameter $0<\beta<1$ to characterize simultaneously the
antigravity strength and AdS curvature:%
\begin{equation}
8\pi\tilde{G}_{N}\equiv3\lambda r_{0}^{2}\frac{\beta^{3}}{1-\beta}%
,\;\;R_{-}=-\frac{12}{r_{0}^{2}}\frac{1-\beta}{\beta^{3}}. \label{beta}%
\end{equation}
All entries in Table 1 are fixed phenomenologically except for $\beta$. Far
from singularities, the antigravity potential $v(h)$ may differ from
$v_{4}(h)$, but similar behavior to Table.1 is expected. This allows $\phi
_{-}^{2}$ to be small (rather than zero) and $h_{-}^{2}$ large (rather than
infinite), as noted in Table 1. From the signs of $R_{\pm}$ and $\Lambda_{\pm
}$, the geometry is asymptotically de Sitter on the gravity side and anti--de
Sitter on the antigravity side. A static, spherically symmetric metric with
these asymptotics is%
\begin{equation}%
\begin{array}
[c]{l}%
ds^{2}=-dt^{2}A\left(  r\right)  +\frac{1}{A\left(  r\right)  }dr^{2}%
+r^{2}d\Omega^{2},\\
A\left(  r\right)  =\left(  1-\frac{r_{0}}{r}-\frac{\Lambda\left(  r\right)
}{3}r^{2}\right)  ,\;-\infty<r<+\infty,\text{ }\\
\Lambda\left(  r\right)  \equiv\theta\left(  r\right)  \Lambda_{+}%
+\theta\left(  -r\right)  \Lambda_{-},\;\;r_{0}\equiv2G_{N}M,
\end{array}
\label{metric2}%
\end{equation}
Here $M$ is the black-hole mass. This geometry is geodesically complete
\cite{BHoles,BST,BSTloop,Sailing}, as discussed below.

Although the above form is justified only asymptotically, we can extend it
over the full range $-\infty<r<+\infty$ by assuming $V_{4}(\phi,s)$ holds
everywhere and taking, $\phi(r)=\theta(r)\phi_{+}+\theta(-r)\phi_{-},$ and
$s(r)=\theta(r)s_{+}+\theta(-r)s_{-}.$ Together with Eq. (\ref{metric2}),
these solve the full equations of motion for $\left(  \phi\left(  r\right)
,s\left(  r\right)  ,g_{\mu\nu}\left(  r\right)  \right)  $, including
derivatives, except at $r=0$. There, the scalars are discontinuous but the
metric is continuous and geodesic completeness of the geometry is preserved.
The derivatives of $(\phi\left(  r\right)  ,s\left(  r\right)  )$ introduce
$\delta(r)$ terms that are unbalanced in the equations of motion only at
$r=0$. This problem can be swept under rug by formally defining the theta
functions $\theta\left(  \pm r\right)  $ to vanish at $r=0$ (consistent with
the red curve in Fig.1). In any case, a continuous, globally defined, analytic
solution$(\phi\left(  r\right)  ,s\left(  r\right)  )$ ---corresponding to the
red curve in Fig.1---is constructed in \cite{IBbhHsolution}. It satisfies the
correct asymptotic limits for $(\phi\left(  r\right)  ,s\left(  r\right)  )$,
significantly modifies the metric from Eq. (\ref{metric2}), and preserves
geodesic completeness. The red curve implies the surprizing re-establishement
of the electroweak symmetry SU$\left(  2\right)  \times$U$\left(  1\right)  $
at the black hole singularity because of the vanishing of the Higgs field
$s\left(  0\right)  =0$.

\section{New geometry in the interior of black holes}

As shown in Fig. 2, the continuous metric function $A(r)$ in (\ref{metric2}$)$
diverges at $r=0$, has zeros at the black-hole horizon $r_{h}$ and
cosmological horizon $r_{c}$, and tends to $\mp\infty$ for large $|r|$ in the
gravity/antigravity regions, respectively. No zeros occur in the antigravity
region. The zeros $(r_{h},r_{c})$ are fixed functions of $(r_{0},\Lambda_{+})$
computed in \cite{IBbhHinterior}. For tiny $\Lambda_{+}$ (as in our universe)
they may be approximated, $r_{h}\simeq r_{0}\left[  1+O(r_{0}^{2}/r_{c}%
^{2})\right]  ,$ and $r_{c}\simeq\sqrt{3/\Lambda_{+}}\simeq\sqrt{8\pi
G_{N}/3\lambda^{\prime}}\simeq1.65\times10^{26}\ \mathrm{m}.$ The visible
spacetime patch, for observes like us, is the region $r_{h}<r<r_{c};$ but for
proper observers that can travel through the horizons, and even singularities
(see below), the full geodesically complete universe is the full range
$-\infty<r<\infty.$%
\begin{figure}[htb]
\centering
        \includegraphics[width=0.7\columnwidth]{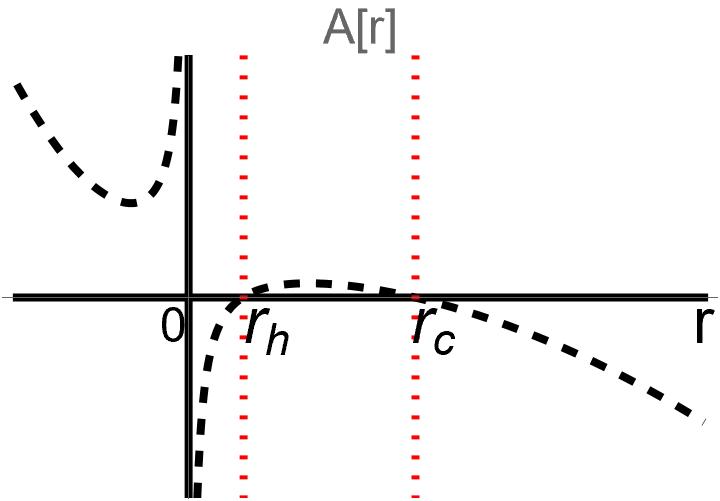}
        \caption{AdSSdS black hole}
        \label{fig:fig2}
\end{figure}
On the gravity side, the geometry is Schwarzschild--de Sitter (SdS);
on the antigravity side, it is a negative-mass
Schwarzschild--anti--de Sitter (to see this, replace
$r\rightarrow-|r|$ for $r<0$). For brevity, I refer to the full
gravity/antigravity spacetime as the \textit{AdSSdS black hole}.

Kruskal--Szekeres (KS) maximal extension of this $\left(  t,r\right)  $
spacetime, in terms of $\left(  u,v\right)  $ coordinates, is given by the
transformation
\begin{equation}%
\begin{array}
[c]{c}%
uv\left(  r\right)  =\left(  e^{r_{\ast}(r)/r_{0}}\right)  ^{\text{Sign}%
\left(  r\right)  }\times\text{Sign}\left(  -rA\left(  r\right)  \right)
,\;\;\\
\frac{v\left(  r\right)  }{u\left(  r\right)  }=e^{t/r_{0}}\times
\text{Sign}\left(  -rA\left(  r\right)  \right),\text{ }-\infty<r<\infty  ,\\
\text{with }r_{\ast}\left(  r\right)  \equiv\int_{0}^{r}\left(
A\left( r^{\prime}\right)  \right)  ^{-1}dr^{\prime}.
\end{array}
\label{uv}%
\end{equation}
Here the symbol $r_{\ast}\left(  r\right)  $ is the well known tortoise
coordinate \cite{tortoise}. The $\mathrm{Sign}(r)$ and $\mathrm{Sign}(-rA(r))$
factors---absent in traditional treatments---are required here to consistently
include the antigravity regions with $r<0$.

Products and ratios of $uv$ and $v/u$ yield $u^{2}$ and $v^{2}$; taking square
roots with appropriate signs give $(u,v)$ for all KS regions ($I-VI$). The
metric in (\ref{metric2}) then depends only on $du\,dv$ and $uv$, agreeing
with \cite{BHoles} in the limit $\Lambda_{\pm}\rightarrow0$ (after adjusting
for the signs of $r$ here versus the strictly positive $r$ in \cite{BHoles}).
\begin{figure}[htb]
    \begin{minipage}{0.48\columnwidth}
        \includegraphics[width=\columnwidth]{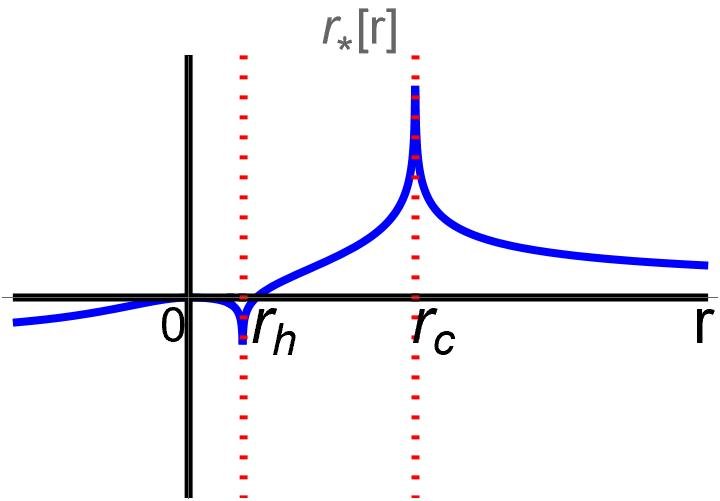}
        \caption{\label{fig:fig3} Tortoise $r_\ast\left(  r\right)$ .}
    \end{minipage}\hfill
    \begin{minipage}{0.48\columnwidth}
        \includegraphics[width=\columnwidth]{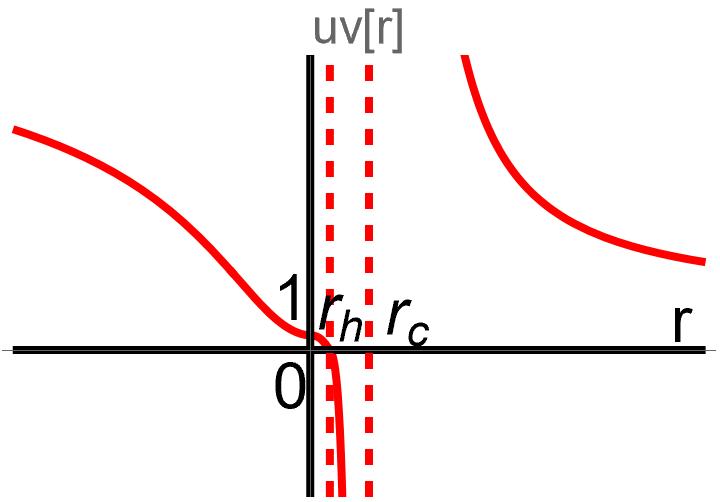}
        \caption{\label{fig:fig4} $uv\left(  r\right)  .$}
    \end{minipage}
\end{figure}
Fig. 3 shows that $r_{\ast}(r)$ diverges at $r_{h},r_{c}$ and vanishes at
$r=0$. Fig. 4 shows that the product $uv=1$ at $r=0$, vanishes at $r_{h}$, is
negative in the visible region $r_{h}<r<r_{c}$, diverges and changes sign at
$r_{c}$, and approaches constants $uv_{\pm\infty}$ in asymptotic limits
$r\rightarrow\pm\infty$. Exact expressions for $uv_{\pm\infty}$ are given in
\cite{IBbhHinterior}; approximations are: $uv_{+\infty}\simeq r_{c}/\left(
r_{h}\sqrt{e}\right)  $ (huge in our universe)$,$ while $uv_{-\infty}%
(\beta)\ $is a specific function of $\beta$ that simplifies in the
neigborhoods of $\beta\rightarrow0,1.$ Namely, for weak antigravity (large AdS
curvature), $uv_{-\infty}\left(  \beta\simeq0\right)  \simeq1+O\left(
\beta^{2}\right)  $ and for strong antigravity (small AdS curvature),
$uv_{-\infty}(\beta\simeq1)\simeq\sqrt{e(1-\beta)}\,\exp\!\left[  \frac{\pi
}{2}/\sqrt{1-\beta}\right]  \ \simeq\ \infty.$

This information about $uv$ suffices to construct the KS diagram for AdSSdS
displayed in \cite{IBbhHinterior}. It contains additional antigravity regions
$(V_{h},VI_{h})$ absent in the traditional case. These lie between the
singularity curve $uv=1$ and the hyperbolic antigravity boundary
$uv\rightarrow uv_{-\infty}(\beta)$ as $r\rightarrow-\infty$. For weak
antigravity, $uv_{-\infty}\left(  \beta\simeq0\right)  \simeq1,$ regions
$(V_{h},VI_{h})$ shrink in size; for strong antigravity, $uv_{-\infty}\left(
\beta\simeq1\right)  \simeq\infty,$ regions $(V_{h},VI_{h})$ expand in size to
fill the entire ($V,VI$) sectors.

\section{Causal structure and flow of information}

The Penrose diagram in Fig. 5, in $\left(  \tilde{u},\tilde{v}\right)  $
coordinates, provides a clearer view of geodesic completeness and causality.
$\left(  \tilde{u},\tilde{v}\right)  $ are related to $(u,v)$ according to:
$u=\tan\tilde{u},\;\;v=\tan\tilde{v}.$ The product $uv$, at critical points in
Fig. 4, yields the following equations for curve segments in the $(\tilde
{u},\tilde{v})$ plane, decribing the region boundaries in Fig. 5:
\begin{equation}%
\begin{array}
[c]{l}%
uv\left(  r\right)  \Rightarrow\left.  \tan\tilde{u}\left(  r\right)
\tan\tilde{v}\left(  r\right)  \right\vert _{r\rightarrow\left(
-\infty,0,r_{h},r_{c},+\infty\right)  }\\
\;\;\;\;=\left(  uv_{-\infty},1,0,\pm\infty,uv_{+\infty}\right)  .
\end{array}
\label{conditions}%
\end{equation}%

\begin{figure}[htb]
\centering
        \includegraphics[width=1\columnwidth]{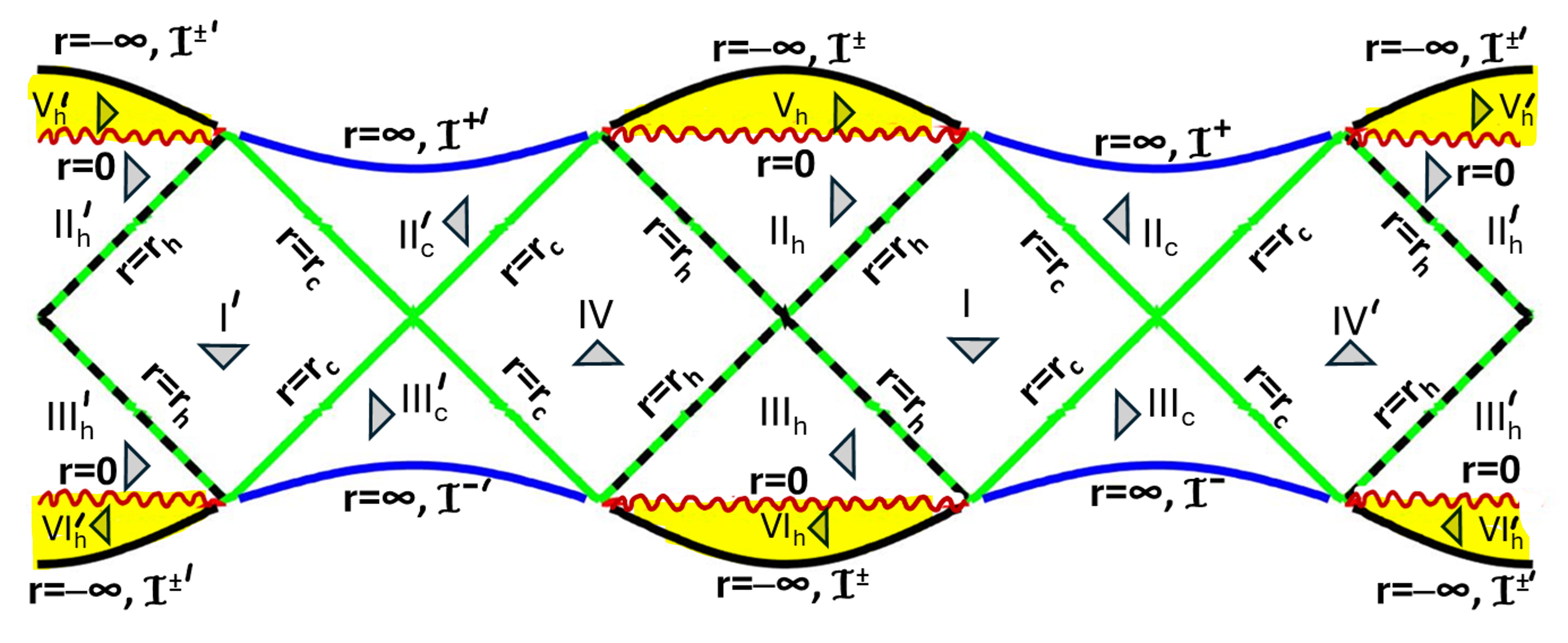}
        \caption{Penrose diagram $\left(  \tilde{u},\tilde{v}\right)  $, complete and
causal.}
        \label{fig:fig5}
\end{figure}

The curve segments that follow from the conditions in
(\ref{conditions}) are interpreted as follows: Black/green dashed
segments: black-hole horizons at $r=r_{h}$; Green segments:
cosmological horizons at $r=r_{c}$; Blue convex curves: asymptotic
boundary of the dS gravity region at $r=+\infty$; Red wavy segments:
black- and white-hole singularities at $r=0$; Black concave curves:
asymptotic boundary of the AdS antigravity region at $r=-\infty$.

The antigravity domains (yellow) lie between the red wavy segments and the
black concave curves and are labelled $(V_{h},V_{h}^{\prime}),\,(VI_{h}%
,VI_{h}^{\prime})$. All white regions correspond to the standard gravity
domains $\left(  I,II_{h},III_{h},IV\right)  $ and $\left(  I^{\prime}%
,II_{c}^{\prime},III_{c}^{\prime},IV^{\prime}\right)  ,$ and they resemble the
Penrose diagram for the SdS black hole given by Gibbons and Hawking
\cite{GibHawk}. The AdSSdS diagram in Fig. 5 is geodesically complete, unlike
the incomplete SdS case in \cite{GibHawk}.

Small right-angled triangles throughout the diagram indicate local forward
lightcones. A massive particle located at the right-angle vertex of a triangle
must propagate within the forward cone; a massless particle travels only along
one of the cone's right-angle edges. These causal rules follow from the
Killing vector $\partial_{t}$ associated with the conserved energy $E$. All
geodesics must obey these propagation rules at each instant of proper time.

\textit{Example}: Consider a massless particle (e.g. photon) in region $I$ at
radial position $r_{0}$ with $r_{h}<r_{0}<r_{c}$, moving radially toward the
black hole with vanishing angular momentum $\vec{L}=0$. It's geodesic (see
Eq.(\ref{tphoton})) is a $(3\pi/4)^{\circ}$ straight line parallel to the
lower $r=r_{h}$ horizon. As proper time increases, the photon crosses into
region $II_{h}$, passes through the upper $r=r_{h}$ horizon, and reaches the
singularity $r=0$ (red wavy line) in a finite amount of proper time, $\tau
_{0}=\frac{r_{0}}{E},$ where $E$ is the photon energy. The geodesic then
continues through the singularity into antigravity region $V_{h}$, requiring
an \textit{infinite} amount of proper time $\tau$ to reach the $r=-\infty$
boundary (black curve). This appears to be a complete geodesic because it is
not artificially truncated at a finite value of proper time, as it would have
happened if the antigravity region had been excised (as in \cite{GibHawk}).

Actually, the geodesic above is not complete yet in the AdSSdS geometry of
Fig.5. Besides the infinite range of proper time $\tau$ which must not be
cutoff artificially, one must also consider what is called the global geometry
of the AdS spacetime \cite{Tong-GR}. According to the global geometry, AdS
spacetime is analogous to a box, such that the $r=-\infty$ boundary marked as
$\mathcal{I}^{\pm}$ in Fig.5. acts like a mirror at the end of the box. So,
although, $\mathcal{I}^{\pm}$ at $r=-\infty,$ is reached in infinite proper
time, it is not the end of global conformal time. Accordingly, the geodesic
discussed above does not end there, it is reflected at the $\mathcal{I}^{\pm}$
boundaries, and then moves downward at an angle of $\left(  \pi/4\right)
^{o}$ following the only permitted causal lightcone direction (the little
triangle) in region $V_{h}.$ So, it continues as a $\left(  \pi/4\right)
^{o}$ straight line that goes through the antigravity region $V_{h},$ sails
through the singularity (wavy line), then the gravity regions $II_{h},$ then
$IV,$ and so on.

Similarly, one can easily figure out the complete geodesics of massless
particles that originate in any region of the AdSSdS Penrose diagram in Fig.5.
This overall picture of complete geodesics alters radically the discussion
concerning the information paradox in black holes; see last section.

\section{Complete geodesics across singularities}

In a spacetime with metric $g_{\mu\nu}(x)$ that is spherically symmetric and
time--translation--invariant---such as (\ref{metric2})---the energy $E$ and
angular momentum $\vec{L}$ are conserved along any geodesic. The geodesic
equations for $(t(\tau),r(\tau))$ as functions of proper time $\tau$ follow
from the effective nonrelativistic constrained Hamiltonian \cite{Hconstraint}:%

\begin{equation}
\left(  \frac{dr}{d\tau}\right)  ^{2}+A\left(  r\right)  \left(  \frac{\vec
{L}^{2}}{r^{2}}+m^{2}\right)  =E^{2},\;\;E=A\left(  r\left(  \tau\right)
\right)  \dot{t}. \label{constr-new}%
\end{equation}

For a massless particle $(m=0)$ with zero angular momentum $(\vec{L}=0)$, Eq.
(\ref{constr-new}) reduces to $r(\tau)=\mp E\tau$, independent of $A(r)$. The
$\pm$ sign corresponds to the propagation direction, consistent with the
Penrose diagram (Fig. 5). The corresponding trajectories are:%

\begin{equation}%
\begin{array}
[c]{l}%
r\left(  \tau\right)  =\mp E\tau,\text{ for any }A\left(  r\right)  ,\text{
iff }m=0=\vec{L},\\
t\left(  \tau\right)  =E\int_{0}^{\tau}d\tau^{\prime}\left(  A\left(  r\left(
\tau^{\prime}\right)  \right)  \right)  ^{-1}\\
\;\;\;\;=\mp\int_{0}^{r\left(  \tau\right)  }dr^{\prime}\left(  A\left(
r^{\prime}\right)  \right)  ^{-1}=\mp r_{\ast}\left(  r\left(  \tau\right)
\right)  .
\end{array}
\label{tphoton}%
\end{equation}
Here $r_{\ast}(r)$ is the tortoise coordinate defined in (\ref{uv}) and
plotted in Fig. 3.

\textit{Example:} Consider the photon in the previous section, starting at
$\tau=-\tau_{0}<0$ in the visible region $I$, with $r_{h}<r_{0}<r_{c}$, moving
inward. For $E>0$, choose $r(\tau)=-E\tau$, describing infall as $\tau$
increases toward $0$. This path matches the complete geodesic described
earlier, traced in the Penrose diagram.

A second solution of (\ref{constr-new}) is $r(\tau)=E|\tau|$, with the same
initial conditions as above. In this case $r\left(  \tau\right)  $ stays
positive for all $-\infty<\tau<\infty$. This geodesic does not cross the
singularity, but it reflects at $r=0$ into region $II_{h}$, then $IV$, and so
forth. This complete geodesic appears naturally in $i(\mathrm{SM}%
+\mathrm{GR})$ but is absent in $(\mathrm{SM}+\mathrm{GR})$, which offers no
prescription for the next instant beyond $\tau=0$ at $r=0$.

\textit{Traversal of the singularity:} Because $r(\tau)$ is independent of
$A(r)$, the singularity in $A(r)$ at $r=0$ does not obstruct passage at
$\tau=0$; only $t(\tau)$ depends on $A(r)$. The divergences of $t(\tau
)=-r_{\ast}(r(\tau))$ at $r=r_{h}$ and $r=r_{c}$ (Fig. 3) explain why
observers in the visible region cannot see beyond these horizons in $t$-time.
Equation (\ref{tphoton}) thus shows that radially infalling massless
particles---photons, gravitons, gluons---can cross both the horizon and the
singularity. After crossing, reaching $r=-\infty$ in the antigravity region
requires infinite proper time. The combined gravity/antigravity spacetime is
therefore geodesically complete for all $-\infty<\tau<\infty$, with the
singularity acting as a traversable bridge for massless particles.

\textit{Massive and nonradial geodesics:} For $m\neq0$ or $L^{2}\neq0$, the
effective potential in (\ref{constr-new}) forms a barrier at the singularity.
Classically, such geodesics may reflect at $r=0$ and follow paths
$II_{h}\rightarrow IV$, remaining geodesically complete. Quantum tunneling
across the barrier into the antigravity region is possible, depending on the
form of $A(r)$.

\textit{Electroweak symmetry restoration at the singularity:} The geodesics of
massive degrees of freedom in the complete theory $i$(SM+GR) are not so simple
as above, since all masses arise from $\phi$ and the Higgs field $s$. In
geodesic calculations, $m$ should be replaced by a dynamical mass
$m(r)=g\,s(r)$, where $g$ is a constant coupling and $s(r)$ is the Higgs
profile \cite{Sailing}. Solving the coupled field equations yields
trajectories $(\phi(r),s(r))$ connecting gravity and antigravity domains (Fig.
1). The remarkable favored solution (red curve in Fig. 1) has both scalars
vanish at $r=0$, with the ratio $h(r)=s(r)/\phi(r)$ satisfying $h(0)=1$. The
vanishing $s(0)=0$ signals restoration of SU(2)$\times$U(1) electroweak
symmetry, making all SM massive particles massless at the singularity. Their
geodesics in the neighborhood of $r=0$ thus may behave like Eq. (\ref{tphoton}%
), allowing continuous propagation across the singularity. However, the
profile of the black hole $A\left(  r\right)  $ also becomes more singular in
the presence of non-constant $\left(  \phi\left(  r\right)  ,s\left(
r\right)  \right)  ,$ so the traversability of massive geodesics is not
immediately obvious and needs further study.

A similar mechanism was discovered in cosmological solutions where $\phi$ and
$s$ vanish at the big bang \cite{BSTloop}\cite{Sailing}. The results in
\cite{IBbhHsolution} extend this behavior to black hole interiors.

In conclusion, in classical physics, information in the form of massless
particles (and perhaps also massive ones) can flow bidirectionally between
gravity and antigravity regions constrained only by the causal structure in
the Penrose diagram (Fig. 5). Moreover, information for massive particles can
also tunnel bidirectionally under the potential barrier (Eq.(\ref{constr-new}%
)) located at the singularity.

\section{Unified gravity--antigravity spacetime}

For simplicity, this paper has modeled black holes as \textit{eternal} black
holes, without truncating their Penrose diagrams to reflect their formation
history (primordial origin, stellar collapse, or mergers). While such details
can be incorporated in future work, they are omitted here to focus on the
essential features of the new complete spacetime structure that includes
antigravity domains.

The discussion so far has centered on a single black hole. Yet in the visible
universe there exists a vast population of black holes, and within each,
$i(\mathrm{SM}+\mathrm{GR})$ predicts an interior antigravity domain. Whether
these interior antigravity regions are interconnected remains an open
question. Regardless, a truly \textit{global}, geodesically complete spacetime
must include not only the traditionally recognized gravity regions inside and
outside black holes, but also all antigravity domains (yellow in Fig.5) inside
every black hole, together with the regions beyond cosmological
horizons---$II_{c},III_{c},II_{c}^{\prime},III_{c}^{\prime}$---and their
periodic repetitions of Fig. 5.

Given such complete geodesics, how is the information puzzle affected? In
conventional SM+GR, and in most current quantum gravity proposals, the
antigravity realms described here are entirely absent. This omission renders
those models geodesically incomplete---a deficiency that directly contributes
to the persistence of the information paradox.

By contrast, $i(\mathrm{SM}+\mathrm{GR})$ predicts that classical information
falling into a black hole is, at least partially---via massless particles such
as gravitons, photons, and gluons---transferred to definite, previously
inaccessible regions of the complete universe. Thus, the question
\textquotedblleft Where does the information go?\textquotedblright\ has a
clear answer in this framework at the level of classical field theory.
Extending this reasoning to string theory level appears feasible along the
lines outlined in \cite{BSTstrings}, thus opening new vistas for quantum gravity.

Importantly, even if some information never returns to the visible
region $I$, unitarity in quantum theory need not be violated. As
long as the complete unified spacetime exists, quantum unitarity
must be defined with respect to \textit{all} observers---those in
region $I$ as well as those in the interior and beyond the
singularity, including those in the antigravity regions. In this
global accounting, information lost to one region is gained by
others, as illustrated in Fig. 5, ensuring overall
unitarity\footnote{Based on certain assumptions, that include
principles such as unitarity and causality, certain S-matrix
arguments show that gravity must always be attractive $G>0$ (see,
for instance, sec 2.2 of \cite{assumptions}). This is in conflict
with the current paper that claims the gravitational strength
$G\left(  x\right)  $ can be negative while both unitarity and
causality (as in the Penrose diagram) remain intact. The resolution
of this conflict must be in the assumptions about the meaning of
unitarity entertained in \cite{assumptions} that possibly does not
include the overall setup of spacetime of the current paper. Namely,
geodesically complete spacetime contains regions of
gravity/antigravity that join at the singularities, while observers
exist everywhere (not only in the gravity sector). It would be of
interest to generalize the arguments in \cite{assumptions}
accordingly.}.

From a quantum information perspective, including ideas such as the ER = EPR
conjecture \cite{EREPR}\cite{ErEPR2}, it is notable that the present work, in
the context of $i(\mathrm{SM}+\mathrm{GR}),$ demonstrates the existence of a
classical communication path between regions $I$ and $IV,$ via the black hole
and the white hole (assuming the latter is real). These channels were absent
in previous discussions of quantum information in the context of black holes.

Moreover, the extensive body of research in AdS/CFT \cite{AdSCFT1}%
-\cite{AdsCFT3} can now be applied to investigate the antigravity interiors of
black holes predicted by $i(\mathrm{SM}+\mathrm{GR})$. In this case the AdS
region is not a toy model since it is a physical region within the black hole.

Are there observable effects outside the horizon that could test
this unified gravity--antigravity picture? Yes, some effects were
discussed in \cite{IBAJ}. Moreover, once  the profiles
$(\phi(r),s(r))$ are determined (Fig. 1), the corresponding metric
$g_{\mu\nu}(r)$ will differ from Eq. (\ref{metric2}), altering
curvature both inside and outside the horizon. These exterior
curvature modifications could, in principle, influence galactic
rotation curves and other gravitational phenomena, such as
gravitational lensing, near black holes. Precise measurements might
allow such extra curvature effects to be distinguished from---or to
supplement---the influence of dark matter. This offers a novel
phenomenological avenue for testing the ideas presented here. Full
details of the computations of $(\phi (r),s(r),g_{\mu\nu}(r))$ will
be given in \cite{IBbhHsolution}.


\begin{thebibliography}{00}                                                                                               %


\bibitem {IBbhHinterior}Itzhak Bars, \textquotedblleft The Higgs field governs
the interior spacetime of black holes,\textquotedblright\ Phys. Rev.
\textbf{D 112} (2025) 7, 075041 [Arxiv:2509.06800 [hep-th]].

\bibitem {MTW}C. Misner, K. Thorne, J. Wheeler, \textquotedblleft%
\textit{Gravitation},\textquotedblright\ see Figure 31.3. W.H.Freeman and Co. [1973].

\bibitem {matrix}C. V. Johnson, \textquotedblleft Wigner meets 't Hooft near
the black hole horizon,\textquotedblright\ Int. J. Mod. Phys. \textbf{D 31}
(2022) 14, 2242003 [Arxiv: 2206.03509 [hep-th] ].

\bibitem {Geometry}J. T. Wheeler, \textquotedblleft Weyl
Geometry,\textquotedblright\ \textit{Gen. Relativ. Gravit.} \textbf{50} (2018)
80, 180103178 [gr-qc].

\bibitem {BST}Itzhak Bars, Paul Steinhardt and Neil Turok, \textquotedblleft
Local conformal symmetry in physics and cosmology,\textquotedblright\ Phys.
Rev. \textbf{D 89} (2014) 043515 [Arxiv: 1307.1848 [hep-th]].

\bibitem {BSTstrings}Itzhak Bars, P. Steinhardt and N. Turok,
\textquotedblleft Dynamical string tension in string theory with
spacetime Weyl invariance,\textquotedblright\ Fortsch. Phys.
\textbf{62} (2014) 901 [Arxiv: 1407.0992 [hep-th]].

\bibitem {IBAJ} Itzhak Bars and Albin James, \textquotedblleft Physical
Interpretation of Antigravity,\textquotedblright\ Phys. Rev.
\textbf{D 93} (2016) 044029 [Arxiv: 1511.05128 [hep-th] ].

\bibitem {BSTloop}Itzhak Bars, Paul Steinhardt and Neil Turok,
\textquotedblleft Antigravity and the Big Crunch/Big Bang
Transition,\textquotedblright\ Phys. Lett. B 715 (2012) 278 [ Arxiv: 1112.2470 [hep-th]].

\bibitem {Sailing}Itzhak Bars, P. Steinhardt and N. Turok, \textquotedblleft
Sailing through the big crunch-big bang transition,\textquotedblright%
\ Phys.Rev.D 89 (2014) 6, 061302 [Arxiv: 1312.0739 [hep-th]].

\bibitem {BHoles}I. J. Araya, Itzhak Bars and A. James, \textquotedblleft
Journey Beyond the Schwarzschild Black Hole Singularity,\textquotedblright%
\ [Arxiv: 1510.03396 [hep-th]].

\bibitem {IBbhHsolution}Itzhak Bars, \textquotedblleft Restoration of
SU(2)$\times$U(1) electroweak symmetry at gravitational
singularities,\textquotedblright\ in preparation.

\bibitem {tortoise}See Eq.(25.31) in reference \cite{MTW}.

\bibitem {GibHawk}G. W. Gibbons and S. W. Hawking, \textquotedblleft
Cosmological event horizons, thermodynamics and particle
creation,\textquotedblright\ Phys. Rev. \textbf{D 15} (1977) 2738. See Fig.4.

\bibitem {Tong-GR}David Tong, \textquotedblleft Lectures on General
Relativity,\textquotedblright\ http://www.damtp.cam.ac.uk/user/tong/gr/gr.pdf.
See Fig.39 on page 179, and related commentary, and Eqs.(4.23,4.25.4.27).

\bibitem {Hconstraint}See Eqs.(25.16a and 25.16b) in reference \cite{MTW}.

\bibitem {assumptions}Simon Caron-Huot, Dalimil Mazac, Leonardo Rastelli,
David Simmons-Duffin, \textquotedblleft AdS Bulk Locality from Sharp
CFT Bounds ,\textquotedblright\ Arxiv: 2106.10274 [hep-th] .

\bibitem {EREPR}J. Maldacena and L. Susskind, \textquotedblleft Cool horizons
for entangled black holes," Fortsch. Phys. \textbf{61} (2013) 781 [Arxiv:
1306.0533 [hep-th]].

\bibitem {ErEPR2}L. Susskind, \textquotedblleft ER=EPR, GHZ, and the
consistency of quantum measurements,\textquotedblright\ [Arxiv: 1412.8483 [hep-th]].

\bibitem {AdSCFT1}J. Maldacena, \textquotedblleft The large N limit of
superconformal field theories and supergravity,\textquotedblright\ Adv. Theor.
Math. Phys. \textbf{2} (1998) 231 [e-print :hep-th/9711200]

\bibitem {AdsCFT2}S. Gubser, I. Klebanov, A. Polyakov, \textquotedblleft Gauge
theory correlators from noncritical string theory,\textquotedblright\ Phys.
Lett. \textbf{B 428} (1998) 105 [e-print: hep-th/9802109].

\bibitem {AdsCFT3}E. Witten, \textquotedblleft Anti de Sitter space and
holography,\textquotedblright\ Adv. Theor. Math. Phys. \textbf{2} (1998) 253
[e-print: hep-th/9802150].
\end{thebibliography}
\end{document}